\documentclass{article}
\usepackage{spconf,amsmath,epsfig}
\usepackage{comment}
\usepackage{multicol}
\usepackage{blindtext,graphicx}
\usepackage[absolute]{textpos}
\usepackage{bm}
\usepackage{multicol}
\usepackage{floatrow}
\usepackage{subcaption}
\usepackage{hyperref}

\graphicspath{{./Images/}}

\title{Explainable Systematic Analysis for Synthetic Aperture Sonar Imagery} 
\name{Sarah Walker$^1$, Joshua Peeples$^1$, Jeff Dale$^2$, James Keller$^2$, Alina Zare$^1$\thanks{This material is based upon work supported by the National Science Foundation Graduate Research Fellowship under Grant No. DGE-1842473 and by the Office of Naval Research grant N00014-16-1-2323. Distribution Statement A: Approved for public release: distribution is unlimited }}
\addressaffone{$^1$Electrical and Computer Engineering, University of Florida, Gainesville, FL, 32611}
\addressafftwo{$^2$Computer Science and Electrical Engineering, University of Missouri, Columbia, MO, 65211}
\begin{document}
%
\maketitle
\begin{abstract}
	 In this work, we present an in-depth and systematic analysis using tools such as local interpretable model-agnostic explanations (LIME) \cite{ribeiro2016should} and divergence measures to analyze what changes lead to improvement in performance in fine tuned models for synthetic aperture sonar (SAS) data. We examine the sensitivity to factors in the fine tuning process such as class imbalance. Our findings show not only an improvement in seafloor texture classification, but also provide greater insight into what features play critical roles in improving performance as well as a knowledge of the importance of balanced data for fine tuning deep learning models for seafloor classification in SAS imagery.
\end{abstract}
\section{Introduction}
\label{sec:intro}

As the field of machine learning expands, its applications have become broader and its performance more critical. New algorithms, faster processors, and larger datasets have been developed out of necessity. Training an artificial neural nework (ANN), especially those with convolutional neural networks (CNNs), can be an involved process. Due to this, using ANNs pretrained on large datasets (\textit{e.g.}, ImageNet) can be more advantageous than training an ANN from scratch. 
However, a pretrained network will be tailored to the data (\textit{e.g.}, RGB) a model was initially trained on, rather than the data in a desired application. 

Previously in the literature, incorporating environmental context seems to improve automatic target recognition \cite{williams2014exploiting,zare2009context,du2015possibilistic}. 
Fine tuning a pretrained network using synthetic aperture sonar (SAS) imagery is an approach to achieve environmental identification. 
Throughout the course of our experiments, we analyze the effects of imbalance on the performance of the CNN. Along with the analysis of the data, we seek to observe the changes in the model quantitatively through information theoretic approaches, such as divergence measures, and qualitatively by explaining decisions made by the model through an interpretable, visual tool. By performing this analysis of the data and model, we propose a systematic study that will improve performance for environmental identification of SAS imagery while also providing insight into the model.


\section{Methods}
\label{sec:format}

	\subsection{Divergence Measures}
	
	After computing the distribution of weights, one can measure the ``distance" between the distribution in the weights of the fine tuned and pretrained models. Divergence measures can be defined to compute the similarity (or dissimilarity) between distributions \cite{principe2010information}. Divergences serve as a way to inversely quantify similarity: the smaller the divergence, the higher the similarity (and vice versa). The divergence measure used in this work is the common Kullback-Leibler (KL) divergence. 
	Given two discrete probability densities, $p(x)$ and $q(x)$, the KL divergence $D_{KL}$
	\cite{principe2010information} is defined as: 
	\begin{equation}
		D_{KL}(p \Vert q) = \sum_{x}^{}p(x)log\left(\frac{p(x)}{q(x)}\right).
		\label{Eqn:KL}
	\end{equation} 
	The two probability densities are estimated using histograms of the weights and biases of the CNNs. Through the divergence measure, we quantify the change in the weights of the model and evaluate the statistical significance. 
	\subsection{Local Interpretable Model-agnostic Explanations (LIME)} \label{LIME}
	Local interpretable model-agnostic explanations (LIME) estimates predictions from any machine learning model by learning local, interpretable models that are understandable to humans \cite{ribeiro2016should,adadi2018peeking}. 
	LIME is designed to balance the trade off between fidelity (trustworthiness of explanation) and interpretability. As a result, the objective function, $\xi(x)$, is comprised of two terms:
	\begin{equation}
		\xi(x) = \mathcal{L}(f,g,\pi_x) + \Omega(g), g \in G \label{Eqn:LIME_Loss}
	\end{equation}
	where $\mathcal{L}$ is a fidelity function, $f$ is the class probabilities produced by a model for a sample $x$, $g$ is a binary vector that is an element of the set of interpretable models ($G$) indicating if a component is interpretable, and $\Omega(g)$ is a measure of complexity (\textit{e.g.}, depth of decision trees \cite{ribeiro2016should}). The goal of this objective is to minimize both terms to promote both local fidelity and interpretabilty. 
	The $\pi_x$ term is a locality measure defined by Equation \ref{Eqn:locality} where $z$ is an instance sampled in the original feature space of $x$:
	\begin{equation}
		\pi_x(z) = exp\left(\cfrac{-D(x,z)^2}{\sigma^2}\right).
		\label{Eqn:locality}
	\end{equation}
	$D$ serves as the distance metric (\textit{e.g.}, cosine, Euclidean) between the two instances, $x$ and $z$. The bandwidth of the kernel, $\sigma$, needs to be chosen. 
	The fidelity function, $\mathcal{L}$, for locally weighted squared loss is shown Equation \ref{Eqn:fidelity}:
	\begin{equation}
		\mathcal{L}(f,g,\pi_x) = \sum_{z,z'\in\mathcal{Z}}^{}\pi_x(z)(f(z)-g(z'))^2
		\label{Eqn:fidelity} 
	\end{equation}
	where $\mathcal{Z}$ is a dataset containing sampled instances $z$ and the original model's outputs $f(z)$ as the labels, $z'$ is a binary indicator of perturbed samples used to generate original representation of instances, and $g(z')$ is selected to be a class of linear models where $g(z') = w_gz'$ . 

	\section{RESULTS AND DISCUSSION}
	\subsection{Datasets and Experimental Setup}\label{Datasets}
\begin{figure}[htb!]
			\begin{subfigure}{.24\textwidth}{
					\includegraphics[width=\textwidth]{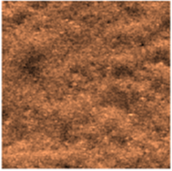}
					\caption{Craters}
					\label{fig:MU_Craters}
				}
			\end{subfigure}
			\begin{subfigure}{.24\textwidth}{
					\includegraphics[width=\textwidth]{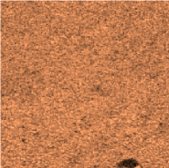}
					\caption{Flat}
					\label{fig:MU_Flat}
				}
			\end{subfigure}
			\begin{subfigure}{.24\textwidth}{
					\includegraphics[width=\textwidth]{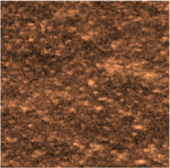}
					\caption{Rocky}
					\label{fig:MU_Rocky}
				}
			\end{subfigure} 
			\begin{subfigure}{.24\textwidth}{
					\includegraphics[width=\textwidth]{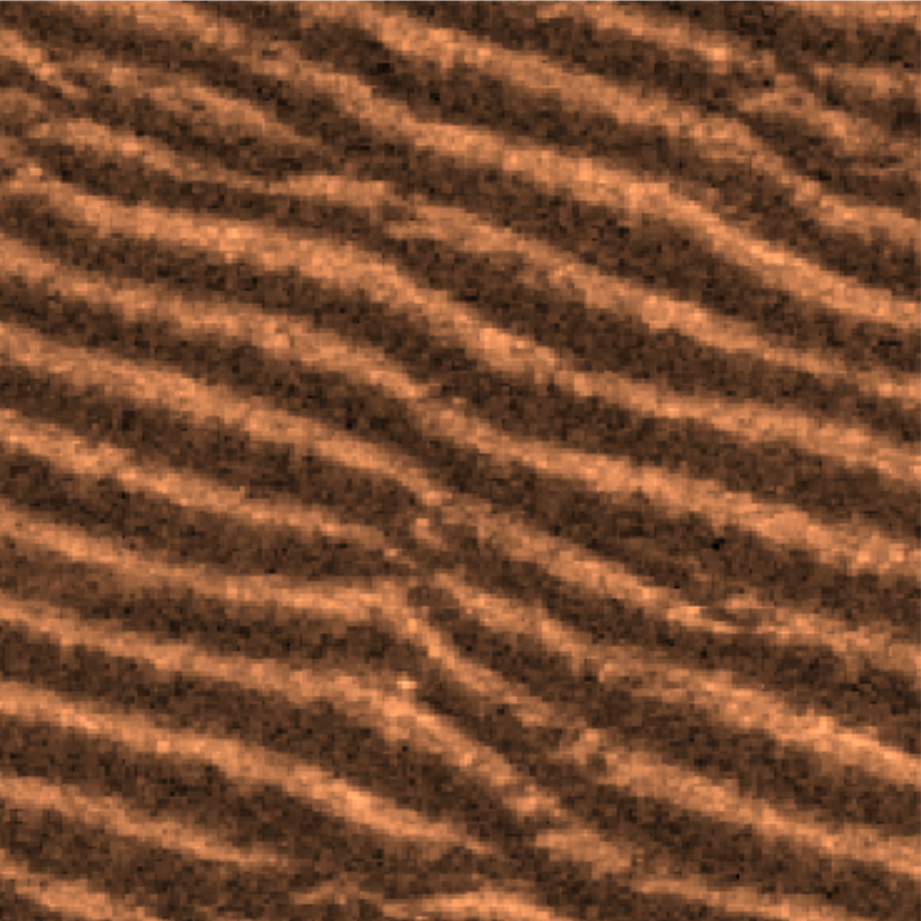}
					\caption{Sand Ripple}
					\label{fig:MU_SR}
				}
			\end{subfigure}
			
			\begin{subfigure}{.24\textwidth}{
					\includegraphics[width=\textwidth]{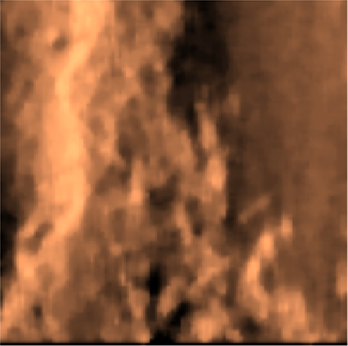}
					\caption{Craters}
					\label{fig:PKNN_Craters}
				}
			\end{subfigure}
			\begin{subfigure}{.24\textwidth}{
					\includegraphics[width=\textwidth]{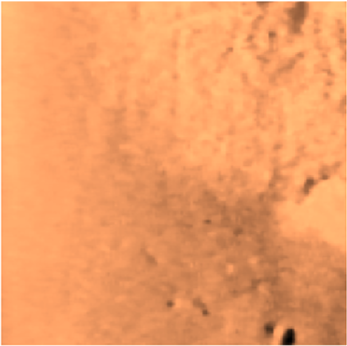}
					\caption{Flat}
					\label{fig:PKNN_Flat}
				}
			\end{subfigure}
			\begin{subfigure}{.24\textwidth}{
					\includegraphics[width=\textwidth]{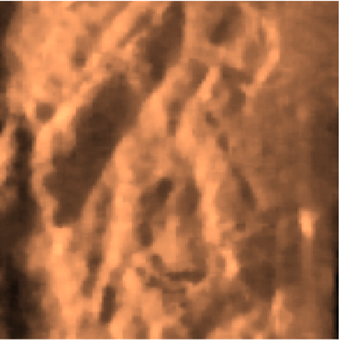}
					\caption{Rocky}
					\label{fig:PKNN_Rocky}
				}
			\end{subfigure} 
			\begin{subfigure}{.24\textwidth}{
					\includegraphics[width=\textwidth]{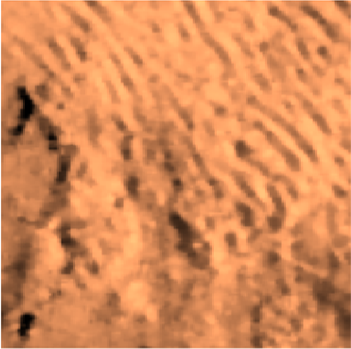}
					\caption{Sand Ripple}
					\label{fig:PKNN_SR}
				}
			\end{subfigure}
		
			\caption{Examples of seafloor texture images from multi site (\ref{fig:MU_Craters}-\ref{fig:MU_SR}) and single site (\ref{fig:PKNN_Craters}-\ref{fig:PKNN_SR}) datasets.}

			\label{fig:MU_examples} 
		\end{figure}

\begin{figure}[htb!]
		\begin{subfigure}{.24\textwidth}{
				\includegraphics[width=\textwidth]{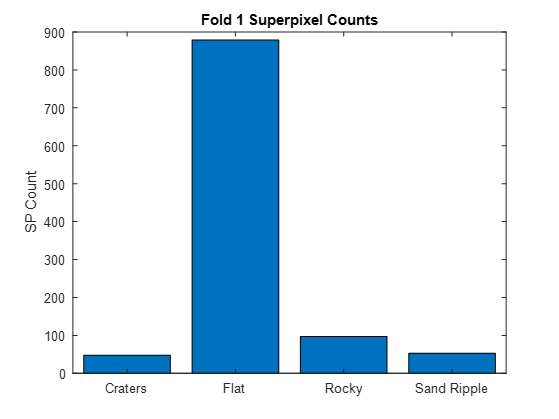}
				\caption{Fold 1}
				\label{fig:PKNN Fold 1}
			}
		\end{subfigure}
		\begin{subfigure}{.24\textwidth}{
				\includegraphics[width=\textwidth]{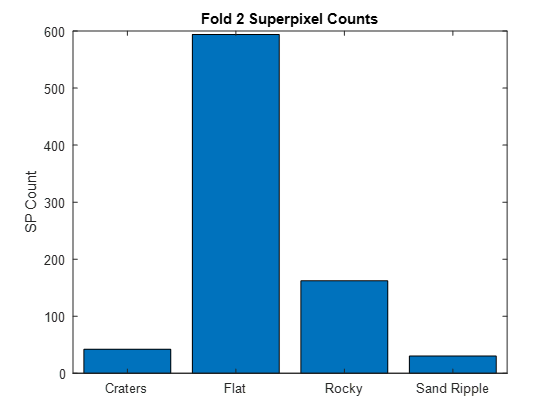}
				\caption{Fold 2}
				\label{fig:PKNN Fold 2}
			}
		\end{subfigure}
		\begin{subfigure}{.24\textwidth}{
				\includegraphics[width=\textwidth]{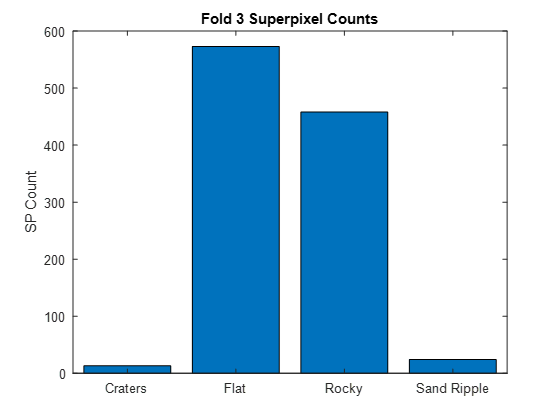}
				\caption{Fold 3}
				\label{fig:PKNN Fold 3}
			}
		\end{subfigure} 
		\begin{subfigure}{.24\textwidth}{
				\includegraphics[width=\textwidth]{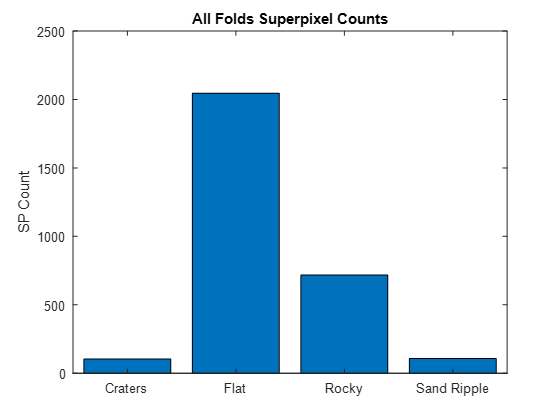}
				\caption{All Folds}
				\label{fig:PKNN_Fold_All}
			}
		\end{subfigure}
		\caption{Distribution of seafloor texture classes for Folds 1 through 3 (\ref{fig:PKNN Fold 1}-\ref{fig:PKNN Fold 3}) and all Folds for single site dataset (\ref{fig:PKNN_Fold_All})}
		\label{fig:PKNN_imbalance}
		\end{figure}

We used two datasets: multi site and single site. The multi site dataset is a combination of SAS images taken from different ocean floor locations whereas the single site dataset is taken at the same location and is composed of superpixels rather than full SAS images. There are a total of 118 and 2,962 images for multi site and single site, respectively. Each dataset contains four human-selected classes displayed in Figure \ref{fig:MU_examples}: flat, rocky, sand ripple, and craters. 
As shown in Figure \ref{fig:PKNN_imbalance}, there is an extreme imbalance in classes for the single site dataset. We used stratified 3-fold cross validation on all experiments resulting in three combinations of testing on a single fold and training with two folds. 
		
We used the cross entropy loss function, Adam optimization, learning rate of 1e-4, batch size of six and 100 epochs to train each CNN. For the baseline model, only the output layer was updated while all other layers' weights are fixed. 
We trained and tested three different classifiers using the 4,096 dimensional features from penultimate layer: $K$-nearest neighbor ($K$NN), support vector machine (SVM), and a fully connected layer (not the same one used for training the model). 
A linear kernel was used for the SVM and the slack variable, $C$, is set to one. The number of nearest neighbors, $K$, was set to three for the $K$NN. For the fully connected layer, we used a linear classifier with the same parameters used during fine tuning, except it was trained with 30 epochs rather than 100. After the model was trained, a softmax was applied to the output for testing. 
	\subsection{Data Characteristics: Balance} \label{sect:data}


We also investigated the impact of class imbalance. To address this problem, 
we used a data sampler \cite{ming2018} to balance the single site dataset.
In Table \ref{tab:PKNN_Balance}, the accuracies for both the balanced and imbalanced datasets are comparable. 
The balanced classifiers trained with fine tuned features achieved slightly better accuracy on average except for the fully connected layer. 
With imbalanced datasets, accuracy is not an optimal measure for performance since the model will benefit from being biased towards the majority class(es). In Table \ref{tab:Crater_PR}, the precision and recall are recorded for one of the minority classes, craters. 
From the precision and recall, the balanced models improve the classification performance of craters for each classifier except for the precision of $K$NN. 
\begin{table}[htb]
	\caption{Test Accuracies of Fine Tuning Experiments for balanced Single Site (Balanced SS), imbalanced Single Site (Imbalanced SS), and the Multi Site (MS)}
	\label{tab:PKNN_Balance}
\begin{tabular}{|c|c|c|c|}
\hline
\multicolumn{4}{|c|}{\textbf{SVM}}                           \\ \hline
\textbf{} & \textit{Balanced SS}   & \textit{Imbalanced SS} & \textit{MS}\\ \hline
\textbf{FT}      & 79.67$\pm$2.19\% & 79.42$\pm$4.37\% & 82.31$\pm$1.88\%         \\ \hline
\textbf{Base}    & 78.94$\pm$2.55\%          & 79.48$\pm$2.02\% & 69.73$\pm$7.05\%         \\ \hline
\multicolumn{4}{|c|}{\textbf{KNN}}                           \\ \hline
\textbf{} & \textit{Balanced SS}   & \textit{Imbalanced SS} & \textit{MS} \\ \hline
\textbf{FT}      & 76.74$\pm$8.61\% & 76.17$\pm$9.20\%4 & 72.17$\pm$4.18\%          \\ \hline
\textbf{Base}    & 72.07$\pm$7.20\%          & 73.44$\pm$10.91\%  & 68.90$\pm$8.94\%          \\ \hline  \multicolumn{4}{|c|}{\textbf{Fully Connected}}               \\ \hline
\textbf{} & \textit{Balanced SS}   & \textit{Imbalanced SS} & \textit{MS}\\ \hline
\textbf{FT}      & 75.19$\pm$7.07\%          & 80.66$\pm$5.11\% & 79.66$\pm$4.20\% \\ \hline
\textbf{Base}    & 70.79$\pm$2.92\%          & 74.09$\pm$9.60\%  & 65.32$\pm$5.53\%        \\ \hline
\end{tabular}
\end{table}
\begin{table*}[htb]
	\centering
	\caption{Single site dataset Crater Precision and Recall for Balance Experiments (Best result bolded for each classifier)}
	\label{tab:Crater_PR}

\begin{tabular}{|c|c|c|c|c|c|c|}
\hline
\textbf{Crater}     & \multicolumn{3}{c|}{\textbf{Precision}}                        & \multicolumn{3}{c|}{\textbf{Recall}}                          \\ \hline
\textbf{Classifier} & \textit{KNN}        & \textit{SVM}        & \textit{FC}        & \textit{KNN}       & \textit{SVM}       & \textit{FC}         \\ \hline
\textbf{Base(B)}    & 10.58$\pm$5.75\%          & 11.67$\pm$6.13\%          & 5.04$\pm$.2.48\%          & \textbf{4.08$\pm$0.95\%} & \textbf{9.42$\pm$1.21\%} &12.36$\pm$3.07\%        \\ \hline          
\textbf{FT(B)}      & 18.52$\pm$13.85\%          & \textbf{15.00$\pm$4.91\%}  &\textbf{8.47$\pm$5.12\%}        &2.21$\pm$1.80\%          & 8.13$\pm$2.97\%          & \textbf{13.75$\pm$5.79\%} \\ \hline   
 \textbf{Base(I)}    & \textbf{20.83$\pm$12.27\%} & 9.59$\pm$2.06\%           & 4.44$\pm$6.29\%           & 2.62$\pm$0.88\%          & 5.05$\pm$1.36\%          & 0.74$\pm$1.04\%       \\ \hline    
\textbf{FT(I)}      & 12.22$\pm$4.16\%          & 14.14$\pm$3.68\%          & 7.41$\pm$6.93\%        & 2.04$\pm$0.48\%         & 5.82.$\pm$2.17\%     & 0.83 $\pm$0.60\%            \\ \hline         
\end{tabular}                          
\end{table*}

\begin{table*}[htb!]
	\caption{KL divergence and Euclidean distance measures for each layer (Largest average difference is bolded for each dataset).}
	\label{tab:Divergences}
	\centering 
\begin{tabular}{|c|c|c|c|c|c|c|}
\hline
\textbf{}          & \multicolumn{3}{c|}{\textbf{$D_{KL}(FT||Base)$}}                                 & \multicolumn{3}{c|}{\textbf{$D_{ED}$}}                      \\ \hline
\textbf{Dataset}   & \textit{Multi}    & \textit{Single}     & \textit{Single(B)}  & \textit{Multi}    & \textit{Single}      & \textit{Single(B)}  \\ \hline
\textbf{1st Conv.} & \textbf{0.148$\pm$0.03} & 0.735$\pm$0.11            & 0.407$\pm$0.16            & 0.109$\pm$0.02  & 0.645$\pm$0.11           & 0.302$\pm$0.12          \\ \hline     \textbf{2nd Conv.} & 0.012$\pm$0.004         & 1.073$\pm$0.38           & 0.305$\pm$0.29            &0.478$\pm$0.15          & 5.291$\pm$1.06            & 2.545$\pm$1.25          \\ \hline
\textbf{3rd Conv.} & 0.007$\pm$0.002         & 1.290$\pm$0.64           & 0.254$\pm$0.27           & 0.847$\pm$0.31          & 8.840$\pm$ 1.83           & 4.259$\pm$2.00           \\ \hline  \textbf{4th Conv.} & 0.006$\pm$0.001         & 2.240$\pm$1.18          & 0.342$\pm$0.40            & 0.967 $\pm$ 0.39          & 10.845$\pm$2.09           & 5.381$\pm$2.47          \\ \hline
\textbf{5th Conv.} & 0.007$\pm$0.003         & 1.456$\pm$0.73           & 0.306$\pm$0.34           & 0.654$\pm$0.31          & 7.674$\pm$1.43           & 4.297$\pm$1.93          \\ \hline
\textbf{1st FC}    & 0.003$\pm$0.004         & \textbf{15.863$\pm$7.89} & \textbf{4.467$\pm$5.04} & \textbf{3.842$\pm$1.77}          & \textbf{44.704$\pm$8.51} & \textbf{27.046$\pm$11.38} \\ \hline
\textbf{2nd FC}    & 0.006$\pm$0.007         & 6.778$\pm$3.56          & 1.286$\pm$1.49          & 3.190$\pm$1.53          & 29.911$\pm$5.78          & 16.519$\pm$6.81          \\ \hline
\end{tabular}
\end{table*}
	
		
	
	\subsection{Model Characteristics: Weights and Explanations}
We also ran experiments to gain insight into the changes in the fine tuned and pretrained models by computing the divergence between the distribution of weights and biases. 
We did not use the weights of the output layer as that was the layer we replaced with our own classifiers. We estimated the probability density functions (PDFs) by computing histograms of the weights. We used Euclidean distances (ED) between the corresponding weights in each layer and KL divergences of the PDFs to measure how the weights of each layer differed.

 In Table \ref{tab:Divergences}, the largest changes occurred in the first convolutional layer of the model for the multi site dataset and the first fully connected layers of the imbalanced and balanced single site. 
 Generally, the later layers in a network should change the most because those features will be more domain specific. However, in the multi site dataset, this is not the case. 
 The SAS imagery is highly textured and these earlier layers will capture features to quantify the texture information in the data. Along with capturing texture features, the first convolution layer is applied at the largest scale (\textit{i.e.}, larger filter sizes) of all layers in the network. 
 As a result, more information will lead to the weights of the first layer needing to adjust the most 
 as opposed to those in the later layers.   
	
For both the balanced and imbalanced single site, each layer had larger changes than the model trained with the multi site dataset. Intuitively, this makes sense because the single site dataset is larger than the multi site dataset. 
Since there are more images in the single site dataset, the later layers are able to update the more domain specific features as well as the earlier descriptors in the network. 
The magnitude of changes for the imbalanced data are larger possibly due to the model's bias towards the majority class (flat) and the model's weights are updating in such a way to favor the largest class. Similar to overfitting, if the weights of the model grow too large, this will hurt the generalization ability of the model to test data that contains samples of both the minority and majority class \cite{buda2018systematic}. 
For the balanced dataset, the divergence measures are smaller in magnitude. This shows that balancing a dataset provides more representative data to train the model, and by indirectly applying regularization to the model by balancing the data, the weights of all the layers of the model will change to capture features beneficial for all classes.
	
\begin{figure}[htb!]
	\begin{subfigure}{.30\textwidth}{
			\includegraphics[width=\textwidth]{Images/MU_SR.png}
			\caption{Sand ripple image}
			\label{fig:MU_LIME_SR}
		}
	\end{subfigure}
	\begin{subfigure}{.30\textwidth}{
			\includegraphics[width=\textwidth]{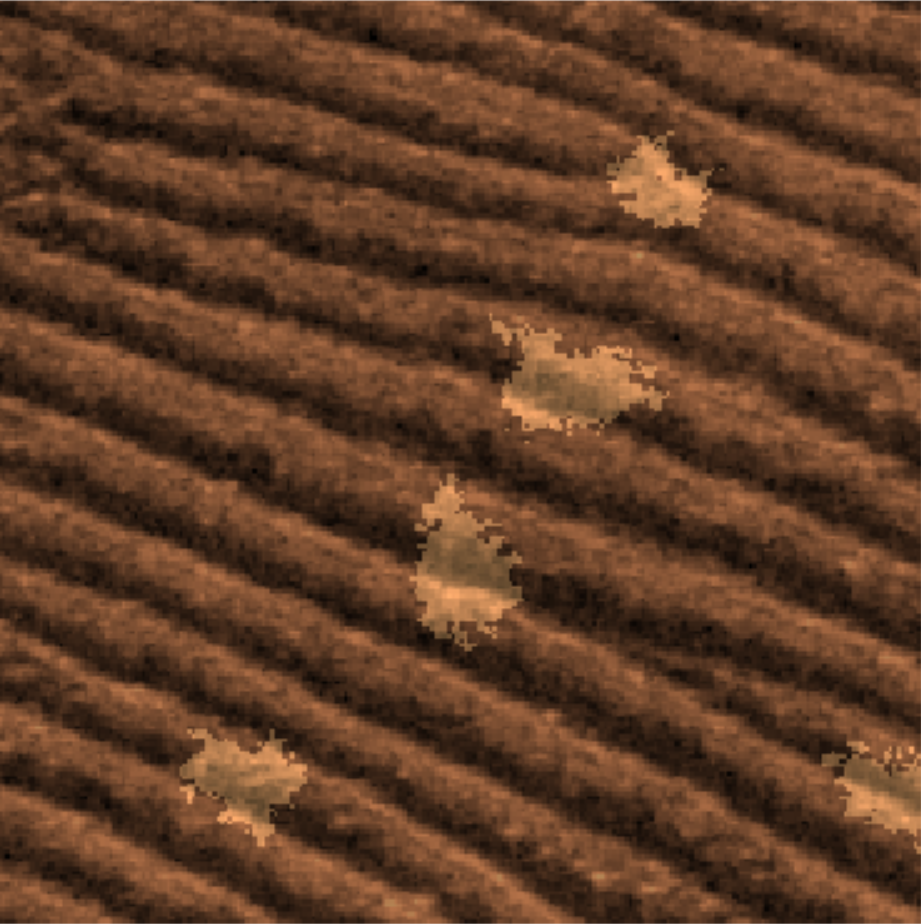}
			\caption{LIME for Baseline AlexNet}
			\label{fig:MU_LIME_Original_SR}
		}
	\end{subfigure}
	\begin{subfigure}{.30\textwidth}{
			\includegraphics[width=\textwidth]{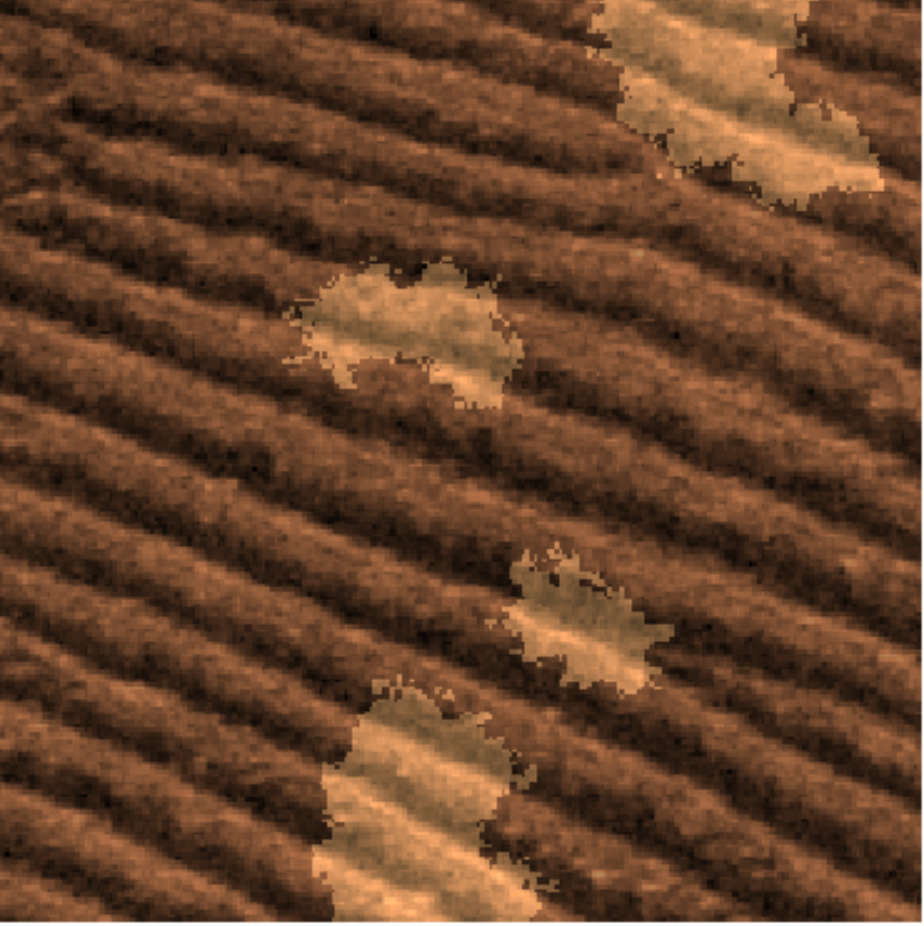}
			\caption{LIME for Fine tuned AlexNet}
			\label{fig:MU_LIME_FT_SR}
		}
	\end{subfigure} 

	\begin{subfigure}{.30\textwidth}{
			\includegraphics[width=\textwidth]{Images/PKNN_SR.png}
			\caption{Sand ripple image}
			\label{fig:PKNN_LIME_SR}
		}
	\end{subfigure}
	\begin{subfigure}{.30\textwidth}{
			\includegraphics[width=\textwidth]{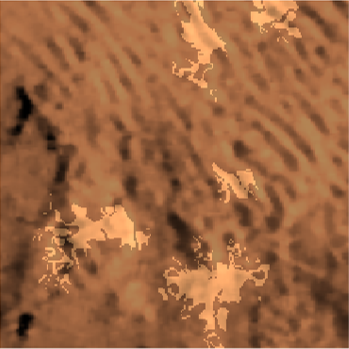}
			\caption{LIME for Baseline AlexNet}
			\label{fig:PKNN_LIME_Original_SR}
		}
	\end{subfigure}
	\begin{subfigure}{.305\textwidth}{
			\includegraphics[width=\textwidth]{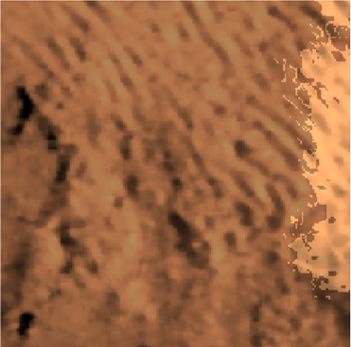}
			\caption{LIME for Fine tuned AlexNet}
			\label{fig:PKNN_LIME_FT_SR}
		}
	\end{subfigure} 
	\caption{LIME visualizations of fine tuned and baseline models for \ref{fig:MU_LIME_SR}-\ref{fig:MU_LIME_FT_SR}) multi site and \ref{fig:PKNN_LIME_SR}-\ref{fig:PKNN_LIME_FT_SR}) balanced single site datasets}
	
	\label{fig:LIME_examples} 
\end{figure}

The results of the LIME experiments can be seen in Figure \ref{fig:LIME_examples}. For the multi site dataset in Figures \ref{fig:MU_LIME_Original_SR} - \ref{fig:MU_LIME_FT_SR}, the baseline model only captures small portions of the image that contain sand ripple. After fine tuning, the model finds more informative pixels that contain larger areas of sand ripple. This result supports the conjecture that scale is important for identifying seafloor textures. 
For the single site dataset, the snippets are not pure texture images as in the multi site dataset. As shown in Figure \ref{fig:PKNN_LIME_Original_SR}, the baseline model captures some sand ripple, but also finds pixels not pertaining to the sand ripple. After fine tuning, as shown in Figure \ref{fig:PKNN_LIME_FT_SR}, the model captures more relevant information as LIME  highlights more sand ripple pixels in larger areas of the image. This is to be expected as fine tuning tailors the model to our data. For texture classification, local regions of the image containing texture(s) of interest are important \cite{ojala2002multiresolution,liu2019bow}. Humans look for similar visual cues in local areas of images to identify patterns such as texture \cite{julesz1962visual,julesz1981textons} and LIME captures similar results. 

	\section{CONCLUSION}
	We presented an approach to analyze deep learning models for SAS imagery identification through classification metrics, a novel divergence measure approach, and qualitative evaluation. 
	Divergence measures provide insight in the differences between the fine tuned and baseline weights. Also, the LIME tool served as a qualitative and interpretable approach to understand the portions of the seafloor imagery that are important once the model is fine tuned. Future work for this analysis includes studying other deep learning models, investigating other sampling approaches coupled with data augmentation techniques such as small rotations and translations (maintaining realistic examples of SAS imagery), and using other explainable models such as saliency maps \cite{adadi2018peeking}.
	Seafloor classification is important for automatic target recognition (ATR) and other applications, but we need to make sure our models are explainable, trustworthy, and reliable before we deploy them in the field. 

\bibliographystyle{IEEEbib}
\bibliography{refs}

\begin{thebibliography}{10}

\bibitem{ribeiro2016should}
Marco~Tulio Ribeiro, Sameer Singh, and Carlos Guestrin,
\newblock ```why should i trust you?' explaining the predictions of any
  classifier,''
\newblock in {\em Proceedings of the 22nd ACM SIGKDD international conference
  on knowledge discovery and data mining}, 2016, pp. 1135--1144.

\bibitem{williams2014exploiting}
David~P Williams and Elias Fakiris,
\newblock ``Exploiting environmental information for improved underwater target
  classification in sonar imagery,''
\newblock {\em IEEE Transactions on Geoscience and Remote Sensing}, vol. 52,
  no. 10, pp. 6284--6297, 2014.

\bibitem{zare2009context}
Alina Zare and Paul Gader,
\newblock ``Context-based endmember detection for hyperspectral imagery,''
\newblock in {\em 2009 First Workshop on Hyperspectral Image and Signal
  Processing: Evolution in Remote Sensing}. IEEE, 2009, pp. 1--4.

\bibitem{du2015possibilistic}
Xiaoxiao Du, Alina Zare, and J~Tory Cobb,
\newblock ``Possibilistic context identification for sas imagery,''
\newblock in {\em Detection and Sensing of Mines, Explosive Objects, and
  Obscured Targets XX}. International Society for Optics and Photonics, 2015,
  vol. 9454, p. 94541I.

\bibitem{principe2010information}
Jose~C Principe,
\newblock {\em Information theoretic learning: Renyi's entropy and kernel
  perspectives},
\newblock Springer Science \& Business Media, 2010.

\bibitem{adadi2018peeking}
Amina Adadi and Mohammed Berrada,
\newblock ``Peeking inside the black-box: A survey on explainable artificial
  intelligence (xai),''
\newblock {\em IEEE Access}, vol. 6, pp. 52138--52160, 2018.

\bibitem{ming2018}
Ming Yang,
\newblock ``Imbalanced dataset sampler,''
  \url{https://github.com/ufoym/imbalanced-dataset-sampler}, 2018.

\bibitem{buda2018systematic}
Mateusz Buda, Atsuto Maki, and Maciej~A Mazurowski,
\newblock ``A systematic study of the class imbalance problem in convolutional
  neural networks,''
\newblock {\em Neural Networks}, vol. 106, pp. 249--259, 2018.

\bibitem{ojala2002multiresolution}
Timo Ojala, Matti Pietikainen, and Topi Maenpaa,
\newblock ``Multiresolution gray-scale and rotation invariant texture
  classification with local binary patterns,''
\newblock {\em IEEE Transactions on pattern analysis and machine intelligence},
  vol. 24, no. 7, pp. 971--987, 2002.

\bibitem{liu2019bow}
Li~Liu, Jie Chen, Paul Fieguth, Guoying Zhao, Rama Chellappa, and Matti
  Pietik{\"a}inen,
\newblock ``From bow to cnn: Two decades of texture representation for texture
  classification,''
\newblock {\em International Journal of Computer Vision}, vol. 127, no. 1, pp.
  74--109, 2019.

\bibitem{julesz1962visual}
Bela Julesz,
\newblock ``Visual pattern discrimination,''
\newblock {\em IRE transactions on Information Theory}, vol. 8, no. 2, pp.
  84--92, 1962.

\bibitem{julesz1981textons}
Bela Julesz,
\newblock ``Textons, the elements of texture perception, and their
  interactions,''
\newblock {\em Nature}, vol. 290, no. 5802, pp. 91--97, 1981.

\end{thebibliography}

\end{document}